\documentclass{iau} 
\usepackage{graphicx}

\title[Starspots and differential rotation in KIC 11560447] 
{Starspot activity and differential rotation in KIC 11560447 }

\author[E. I\c{s}{\i}k et al.]   
{Emre I\c{s}\i k$^{1,2},$
\.{I}brahim \"Ozavc{\i}$^3$, Hakan V. \c{S}enavc{\i}$^3$, Gaitee A.J. Hussain$^{4,5}$, Douglas O'Neal$^6$, Mesut Y{\i}lmaz$^3$, \and Selim O. Selam$^3$}
\affiliation{$^1$Max-Planck-Institut f\"ur Sonnensystemforschung, Justus-von-Liebig-Weg 
3, 37077, G\"ottingen, Germany \\ email: {\tt isik@mps.mpg.de} \\[\affilskip]
$^2$Feza G\"ursey Center for Physics and Mathematics, Bo\u{g}azi\c{c}i 
University, 34684 Istanbul, Turkey \\[\affilskip]
$^3$Dept. of Astronomy and Space Sciences, Ankara University,
Tando\u{g}an 06100 Ankara, Turkey \\[\affilskip]
$^4$European Southern Observatory, Karl-Schwarzschild-Str. 2, 85748 Garching, 
Germany \\[\affilskip]
$^5$Universit\'{e} de Toulouse, UPS-OMP, IRAP, 14 Avenue E. Belin, Toulouse, 31400, France \\[\affilskip]
$^6$Keystone College, School of Arts and Sciences, La Plume, PA, 18440, USA}

\pubyear{2018}
\volume{340}  
\jname{Long-Term Datasets for the Understanding of Solar and Stellar Magnetic Cycles}
\editors{D. Banerjee, J. Jiang, K. Kusano \& S.K. Solanki, eds.}
\begin{document}

\maketitle

\begin{abstract}
Using high-precision photometry from the \textit{Kepler} mission, we investigate patterns of spot activity on the K1-type subgiant component of KIC 11560447, a short-period late-type eclipsing binary. We tested the validity of maximum entropy reconstructions of starspots by numerical simulations. Our procedure successfully captures up to three large spot clusters migrating in longitude. We suggest a way to measure a lower limit for stellar differential rotation, using slopes of spot patterns in the reconstructed time-longitude diagram. We find solar-like differential rotation and recurrent spot activity with a long-term trend towards a dominant axisymmetric spot distribution during the period of observations.
\end{abstract}

\firstsection 
\section{Introduction}
Using uninterrupted, high-precision \textit{Kepler} photometry, we have 
been able to uncover the longitudinal distribution of differentially 
rotating starspots on the primary component of the eclipsing binary 
KIC~11560447 (K1IV+M5V; $P_{\rm orb}\simeq 0.53$~d; $i=88^\circ$). 
We present a summary of the main results of our extensive study of 
this system \cite{ozavci18}. 

\section{Results}
Following stitching and corrections of \textit{Kepler} data, 
we solved the system parameters using simultaneously 
(a) a single-spotted model of the most symmetric 
light curve in hand, and (b) the radial velocities derived from 
high-resolution spectra (see \cite[{\"O}zavc{\i} \etal\ (2018)]{ozavci18} 
for details). 

We used the {\tt DoTS} code \cite{cameron97}, which uses maximum 
entropy regularisation techniuque, 
to invert the light curves for consecutive orbital revolutions. 
As performance tests, we produced multiple series of synthetic spotted 
stars and their light curves, using the system parameters we found. 
Inverting these variable light curves for a thousand consecutive orbits, 
we reached the following results: (1) our procedure detects spontaneous 
conglomerations of unresolved small-scale spots as monolithic large spots;
(2) it successfully tracks the longitudinal migration of up to three strong 
spot clusters superposed on randomly changing background of randomly 
scattered small-scale spots. 

\begin{figure}
\begin{center}
\includegraphics[width=.62\columnwidth]{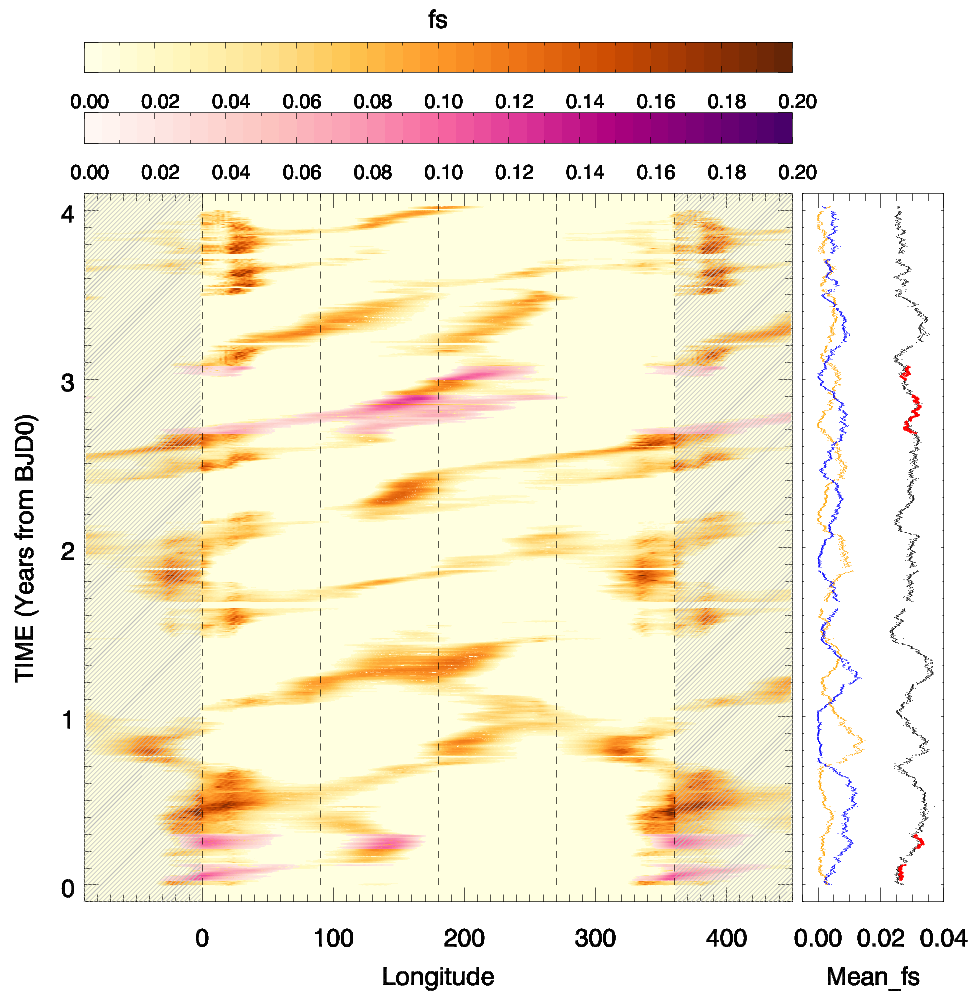} 
\end{center}
 \caption{Time-longitude map of the latitudinally averaged relative spot occupancy 
 on the primary component (orange is for long-, magenta is for short-cadence data). 
 The right panel shows mean values averaged over opposite longitudinal hemispheres 
 (blue/orange) and the global means (black). }
   \label{fig:fs}
\end{figure}

Next, we inverted $\sim 2800$ light curves of the system, using 4-years of 
\textit{Kepler} data, where we stopped the iterations at an optimal $\chi^2$, below 
which the 
procedure attempts to fit photometric noise. The resulting time-longitude diagram 
of the latitudinally averaged relative spot filling factor of the primary component 
is shown in Fig.~\ref{fig:fs}, along with global and hemispheric (east and west) mean 
relative spot coverages. Here, we see that spots or spot groups emerge at certain 
longitudes, and then they decay, while they rotate in most cases faster than the 
orbital rate, hence drift prograde. The hemispheric means show the so-called flip-flop 
pattern. Our numerical experiments have shown that this can be purely related to 
differential rotation. By fitting lines to such features, we were able to 
constrain surface differential rotation, which turned out to be compatible with 
what we get from periodograms of light curves with eclipse effects eliminated. 
We then measured peak-height ratios in periodograms and found that the differential 
rotation is solar-like \cite{ra15}. 
We also found that rotation periods of the prominent spot groups gradually decrease 
within four years, while the system gets fainter. 

\section*{Acknowledgements}
HVS acknowledges support by the Scientific and Technological Research Council of Turkey (T\"{U}B\.{I}TAK) through the project grant 115F033. EI thanks support by the Young Scientist Award Programme BAGEP-2016 of the Science Academy, Turkey.

\end{document}